\documentclass[12pt,preprint]{aastex}
\usepackage{lscape}

\newcommand{\bdv}[1]{\mbox{\boldmath$#1$}}

\def\rel{{\rm rel}}
\def\e{{\rm E}}
\def\au{{\rm AU}} 
\def\muas{{\mu\rm as}}
\def\kms{{\rm km}\,{\rm s}^{-1}}
\def\kpc{{\rm kpc}}
\def\pc{{\rm pc}}
\def\rel{{\rm rel}}
\def\max{{\rm max}}
\def\mas{{\rm mas}}
\def\geo{{\rm geo}}
\def\hel{{\rm hel}}
\def\masyr{{\rm mas\,yr^{-1}}}

\def\e{{\rm E}}
\def\bpi{{\bdv{\pi}}}

\def\bmu{{\bdv{\mu}}}

\def\hel{{\rm hel}}

\begin{document}
\title{The Extreme Microlensing Event OGLE-2007-BLG-224:  Terrestrial
Parallax Observation of a Thick-Disk Brown Dwarf}

\author{
A.~Gould\altaffilmark{1,2,3},
A.~Udalski\altaffilmark{4,5},
B.~Monard\altaffilmark{1,6},
K.~Horne\altaffilmark{7,8,9},
Subo Dong\altaffilmark{1,2},
N.~Miyake\altaffilmark{10,11}, 
K.~Sahu\altaffilmark{8,12},
D.P.~Bennett\altaffilmark{10,8,13}\\
and\\
{\L}.~Wyrzykowski\altaffilmark{14},
I.~Soszy{\'n}ski\altaffilmark{5},
M.K.~Szyma{\'n}ski\altaffilmark{5},
M.~Kubiak\altaffilmark{5},
G.~Pietrzy{\'n}ski\altaffilmark{5,15},
O.~Szewczyk\altaffilmark{15},
{K}.~Ulaczyk\altaffilmark{5}\\
(The OGLE Collaboration)\\
W.~Allen\altaffilmark{16},
G.W.~Christie\altaffilmark{17},
D.L.~DePoy\altaffilmark{18},
B.S.~Gaudi\altaffilmark{2},
C.~Han\altaffilmark{19},
C.-U.~Lee\altaffilmark{20},
J.~McCormick\altaffilmark{21},
T.~Natusch\altaffilmark{22},
B.-G.~Park\altaffilmark{20},
R.W.~Pogge\altaffilmark{2}\\
(The $\mu$FUN Collaboration),\\
A.~Allan\altaffilmark{23}
M.F.~Bode\altaffilmark{24}
D.M.~Bramich\altaffilmark{8,25},
M.J.~Burgdorf\altaffilmark{26},
M.~Dominik\altaffilmark{9,27},
S.N.~Fraser\altaffilmark{24}, 
E.~Kerins\altaffilmark{28}, 
C.~Mottram\altaffilmark{24}, 
C.~Snodgrass\altaffilmark{29},
I.A.~Steele\altaffilmark{24},
R.~Street\altaffilmark{30},
Y.~Tsapras\altaffilmark{30}\\
(The RoboNet Collaboration)\\
F.~Abe\altaffilmark{11}, 
I.A.~Bond\altaffilmark{31}, 
C.S.~Botzler\altaffilmark{32}
A.~Fukui\altaffilmark{11}, 
K.~Furusawa\altaffilmark{11}, 
J.B.~Hearnshaw\altaffilmark{33}, 
Y.~Itow\altaffilmark{11}, 
K.~Kamiya\altaffilmark{11}, 
P.M.~Kilmartin\altaffilmark{34}, 
A.~Korpela\altaffilmark{35}, 
W.~Lin\altaffilmark{31},
C.H.~Ling\altaffilmark{31}
K.~Masuda\altaffilmark{11}, 
Y.~Matsubara\altaffilmark{11}, 
Y.~Muraki\altaffilmark{36}, 
M.~Nagaya\altaffilmark{11}, 
K.~Ohnishi\altaffilmark{37}, 
T.~Okumura\altaffilmark{11},
Y.C.~Perrott\altaffilmark{32}
N.~Rattenbury\altaffilmark{28}, 
To.~Saito\altaffilmark{38}, 
T.~Sako\altaffilmark{11},
L.~Skuljan\altaffilmark{31},
D.J.~Sullivan\altaffilmark{35}, 
T.~Sumi\altaffilmark{11}, 
W.L.~Sweatman\altaffilmark{31},
P.,J.~Tristram\altaffilmark{34}, 
P.C.M.~Yock\altaffilmark{32}\\
({The MOA Collaboration})\\
M.~Albrow\altaffilmark{33} 
J.P.~Beaulieu\altaffilmark{3,39} 
C.~Coutures\altaffilmark{3} 
H.~Calitz\altaffilmark{40} 
J.~Caldwell\altaffilmark{41}
P.~Fouque\altaffilmark{42}
R.~Martin\altaffilmark{43}
A.~Williams\altaffilmark{43}\\
({The PLANET Collaboration})\\
}
\altaffiltext{1}
{Microlensing Follow Up Network ($\mu$FUN)}
\altaffiltext{2}
{Department of Astronomy, Ohio State University,
140 W.\ 18th Ave., Columbus, OH 43210, USA; 
dong,gaudi,gould,pogge@astronomy.ohio-state.edu}
\altaffiltext{3}
{Institut d'Astrophysique de Paris UMR7095, 98bis Boulevard Arago, 75014,
Paris, France.  beaulieu,coutures@iap.fr}
\altaffiltext{4}
{Optical Gravitational Lens Experiment (OGLE)}
\altaffiltext{5}
{Warsaw University Observatory, Al.~Ujazdowskie~4, 00-478~Warszawa,Poland; 
udalski,soszynsk,msz,mk,pietrzyn,szewczyk,kulaczyk@astrouw.edu.pl}
\altaffiltext{6}
{Bronberg Observatory, Centre for Backyard Astrophysics Pretoria, South
Africa, lagmonar@nmisa.org}
\altaffiltext{7}
{RoboNet Collaboration}
\altaffiltext{8}
{Probing Lensing Anomalies NETwork (PLANET) Collaboration}
\altaffiltext{9}
{SUPA, Physics \& Astronomy, North Haugh,
 St Andrews, KY16~9SS, UK; md35,kdh1@st-andrews.ac.uk}
\altaffiltext{10}
{Microlensing Observations in Astrophysics (MOA) Collaboration}
\altaffiltext{11}
{Solar-Terrestrial Environment Laboratory, Nagoya University, 
Nagoya, 464-8601, Japan.}
\altaffiltext{12}
{Space Telescope Science Institute, 3700 San Martin Drive,
Baltimore, MD 21218, USA, ksahu@stsci.edu}
\altaffiltext{13}
{Department of Physics, Notre Dame University, Notre Dame, IN 46556, USA;
bennett@nd.edu}
\altaffiltext{14} {Institute of Astronomy  Cambridge University,
Madingley Rd., CB3 0HA Cambridge, UK wyrzykow@ast.cam.ac.uk}
\altaffiltext{15}{Universidad de Concepci{\'o}n, Departamento de Fisica,
Casilla 160--C, Concepci{\'o}n, Chile}
\altaffiltext{16}
{Vintage Lane Observatory, Blenheim, New Zealand, whallen@xtra.co.nz}
\altaffiltext{17}
{Auckland Observatory, Auckland, New Zealand, gwchristie@christie.org.nz}
\altaffiltext{18}
{Dept.\ of Physics, Texas A\&M University, College Station, TX, USA, 
depoy@physics.tamu.edu}
\altaffiltext{19}
{Department of Physics, Institute for Basic Science Research,
Chungbuk National University, Chongju 361-763, Korea;
cheongho@astroph.chungbuk.ac.kr}
\altaffiltext{20}
{Korea Astronomy and
Space Science Institute, Daejon 305-348, Korea; leecu,bgpark@kasi.re.kr}
\altaffiltext{21}
{Farm Cove Observatory, Centre for Backyard Astrophysics,
Pakuranga, Auckland New Zealand; farmcoveobs@xtra.co.nz}
\altaffiltext{22}
{AUT University, Auckland, New Zealand. tim.natusch@aut.ac.nz}
\altaffiltext{23}
{School of Physics, Univ.\ of Exeter, Stocker Road, Exeter EX4 4QL, UK.
aa@astro.ex.ac.uk}
\altaffiltext{24}
{Astrophysics Research Institute, Liverpool John Moores Univ.,
Twelve Quays House, Egerton Wharf, Birkenhead CH41 1LD, UK;
cjm,snf,ias,mfb,mjb@staru1.livjm.ac.uk}
\altaffiltext{25}
{Isaac Newton Group of Telescopes, Apartado de Correos 321, E-38700 Santa
Cruz de la Palma, Canary Islands, Spain. dmb@ing.iac.es}
\altaffiltext{26}
{Deutsches SOFIA Institut, Universitaet Stuttgart, Pfaffenwaldring 31,
70569 Stuttgart, Germany. mburgdorf@sofia.usra.edu}
\altaffiltext{27}
{Royal Society University Research Fellow}
\altaffiltext{28}
{Jodrell Bank Centre for Astrophysics, Univ.\ of Manchester, Manchester, 
M13 9PL,UK. Eamonn.Kerins@manchester.ac.uk,njr@jb.man.ac.uk}
\altaffiltext{29}
{European Southern Observatory, Alonso de Cordova 3107, Casilla 19001,
Vitacura, Santiago 19, Chile. csnodgra@eso.org}
\altaffiltext{30}
{Las Cumbres Observatory Global Telescope Network, 6740B Cortona Dr, 
Suite 102, Goleta, CA, 93117, USA. rstreet,ytsapras@lcogt.net}
\altaffiltext{31}
{Institute of Information and Mathematical Sciences, Massey University,
Private Bag 102-904, North Shore Mail Centre, Auckland, New Zealand;
i.a.bond,w.lin,c.h.ling,l.skuljan,w.sweatman@massey.ac.nz}
\altaffiltext{32}
{Department of Physics, University of Auckland, 
Private Bag 92019, Auckland, New Zealand;
c.botzler,p.yock@auckland.ac.nz,yper006@aucklanduni.ac.nz}
\altaffiltext{33}{University of Canterbury, Department of Physics and 
Astronomy, Private Bag 4800, Christchurch 8020, New Zealand.}
\altaffiltext{34}
{Mt.\ John Observatory, P.O. Box 56, Lake Tekapo 8770, New Zealand.}
\altaffiltext{35}
{School of Chemical and Physical Sciences, Victoria University, 
Wellington, New Zealand.a.korpela@niwa.co.nz,denis.sullivan@vuw.ac.nz}
\altaffiltext{36}
{Department of Physics, Konan University, Nishiokamoto 8-9-1, 
Kobe 658-8501, Japan.}
\altaffiltext{37}
{Nagano National College of Technology, Nagano 381-8550, Japan.}
\altaffiltext{38}
{Tokyo Metropolitan College of Industrial Technology, Tokyo 116-8523, Japan.}
\altaffiltext{39}
{Univ.\ College of London, Department of Physics and Astronomy, Gower
Street, London WC1E 6BT, UK.}
\altaffiltext{40}
{Dept.\ of Physics/Boyden Observatory, University of the Free State,
Bloemfontein 9300, South Africa. calitzjj.sci@mail.uovs.ac.za}
\altaffiltext{41}
{McDonald Observatory, 16120 St Hwy Spur 78 \#2, Fort Davis, TX 79734, USA.
caldwell@astro.as.utexas.edu}
\altaffiltext{42}
{Laboratoire d'Astrophysique (UMR 5572), Univ.\ Paul Sabatier - Toulouse
3, 14, avenue Edouard Belin, F-31400 Toulouse, France. pfouque@ast.obs-mip.fr}
\altaffiltext{43}
{Perth Observatory, Walnut Road, Bickley, Perth 6076, Australia.
rmartin,andrew@physics.uwa.edu.au}

\begin{abstract}

Parallax is the most fundamental technique to measure distances to
astronomical objects.  Although terrestrial parallax was pioneered
over 2000 years ago by Hipparchus (ca.\ 140 B.C.E.)
to measure the distance to the Moon,
the baseline of the Earth is so small that terrestrial parallax 
can generally only be
applied to objects in the Solar System.  However, there exists a class
of extreme gravitational microlensing events in which the effects of
terrestrial parallax can be readily detected and so permit the
measurement of the distance, mass, and transverse
velocity of the lens.  Here we
report observations of the first such extreme microlensing event
OGLE-2007-BLG-224, from which we infer that the lens is a brown dwarf of
mass $M=0.056\pm 0.004\,M_\odot$, with a distance of $525\pm 40\,\pc$ 
and a transverse
velocity of $113\pm 21\,\kms$.  The velocity places the lens
in the thick disk, making this the lowest-mass thick-disk
brown dwarf detected so far.  
Follow-up observations may allow one to observe the
light from the brown dwarf itself, thus serving as an important
constraint for evolutionary models of these objects and potentially
opening a new window on sub-stellar objects.  The low a priori
probability of detecting a thick-disk brown dwarf in this event, when
combined with additional evidence from other observations, suggests that
old substellar objects may be more common than previously assumed.

\end{abstract}


\section{Discovery of an Extreme Microlensing Event (EME)
\label{sec:intro}}

By several measures OGLE-2007-BLG-224 was the most
extreme microlensing event (EME) ever observed, having a substantially
higher magnification, shorter-duration peak, and faster angular speed
across the sky than any previous well-observed event.
These extreme features suggest an extreme lens.
Fortunately, the observations themselves had extreme characteristics,
and these permit one to test the nature of the lens.

OGLE-2007-BLG-224 (RA=18:05:41, Dec=$-28$:45:36)
was announced as a probable microlensing event
by the OGLE collaboration \citep{ews} on 9 May 2007 and independently by
the MOA collaboration \citep{bond02} on 12 May
as MOA-2007-BLG-163.  At discovery, it was
already recognized to be quite short, Einstein timescale $t_\e  \sim 7\,$days,
and was consistent with peaking at high magnification, although
with very low probability of doing so.  Such high-mag
events (typically defined as $A_\max>100$)
are extremely sensitive to planets, and hence planet-detection
groups attempt to predict these rare
events so as to intensively monitor them over their peak. This
is notoriously difficult, especially for short events,
because so little information is available before peak.  Indeed,
four of the five high-mag events that have to date yielded planet detections
had much longer-than-average Einstein timescales, $t_\e>40\,$days
\citep{ob05071,ob05169,ob06109,mb07192,mb07400}.
Nevertheless, 20 hours before peak, OGLE issued an alert calling this
a ``possible high-magnification event'', and 10 hours later,
by combining OGLE and MOA data, including points from MOA
just 12 hours before peak, the $\mu$FUN collaboration
was able to predict $A_\max>50$
and on this basis issued a general alert advocating intensive
observations.  

As a result, on the night of the peak,
the $\mu$FUN Bronberg 0.35m telescope (South Africa) obtained 754 
unfiltered images over 6.5 hours ending at UT 03:43 just 10 minutes
before peak, during which the magnified flux increased by a
factor 38.  The $\mu$FUN SMARTS 1.3 m telescope (Chile) obtained
a total of 304 images (52 in $I$ band, 8 in $V$, and 244 in $H$)
over 5.5 hours beginning 38 minutes before peak; and the OGLE 1.3m Warsaw
telescope obtained 62 images (57 in $I$ and 5 in $V$)
over 6.9 hours beginning 25 minutes before peak. The RoboNet 2m Liverpool
Telescope obtained 8 $R$ images whose number and timing were determined
by an automated program \citep{snodgrass08,tsapras09}
that queries several photometry databases and
attempts to optimize observations.  Critically, two of these observations
were about 35 minutes before peak and two others were 25 minutes after peak.

Additional observations were obtained by the MOA 1.8m telescope (New Zealand),
the $\mu$FUN 0.4m Auckland, 0.35m Farm Cove, and 0.4m Vintage Lane
telescopes (New Zealand), the $\mu$FUN 1.0m Mt.\ Lemon telescope (Arizona),
the PLANET 1.0m Canopus (Tasmania), 0.6m (Perth), 1.5m Boyden (South Africa)
and 1.54m Danish (Chile) telescopes, as well as the RoboNet Faulkes North 2m
(Hawaii), and Faulkes South 2m (New South Wales) telescopes.  These
observations helped in the determinations of the event's overall 
parameters, but
not in the characterization of the peak, which is the central focus
of this {\it Letter}.

\section{EME Permits Measurement of Two Higher-Order Parameters
\label{sec:higher-order}}

Normally, microlensing events yield only one physical characteristic parameter,
the Einstein timescale $t_\e$, which is a degenerate combination of
three physical parameters of the system, i.e., the lens mass $M$, the
lens-source relative proper motion in the Earth frame, 
$\bmu_\geo$, and the lens-source relative parallax $\pi_\rel=\au/D_l - \au/D_s$,
\begin{equation}
t_\e = {\theta_\e\over \mu_\geo}; \quad \theta_\e = {\pi_\rel\over \pi_\e};
\quad \pi_\e^2 = {\pi_\rel\over\kappa M},
\label{eqn:thetae}
\end{equation}
where $D_l$ and $D_s$ are the lens and source distances and 
$\kappa = 4 G /(c^2\au) \sim 8.1\,\mas\,M_\odot^{-1}$.   Here, 
$\pi_\e=\au/\tilde r_\e$ is the
``microlens parallax'', the ratio of the radius of the Earth's orbit
to the Einstein radius projected on the observer plane, $\tilde r_\e$.
See Figure 1 of \citealt{gould00} for a geometric derivation of the
relations between $(\pi_\e,\theta_\e)$ and $(M,\pi_\rel)$.

Hence, if the lens is unseen, its mass and distance can be determined
only by measuring both $\theta_\e$ and $\pi_\e$ \citep{gould92}.
It is quite rare that
either of these parameters can be measured, and only three times (out of
$>4000$ events) have both been well-measured without seeing the lens itself
\citep{an02,jaros05,kubas05}.  
However, as pointed out more
than a decade ago, EMEs present a unique 
opportunity to measure both parameters \citep{gould97}.

\subsection{Angular Einstein Radius, $\theta_\e$, From Finite-Source Effects
\label{sec:thetae}}

First, during an EME peak, the lens is very likely to transit the
source, giving rise to distinctive ``finite-source'' effects that
yield $\rho = \theta_*/\theta_\e$, the ratio of the source radius to
the Einstein radius.  At high magnification, $A=1/u$, where $u$ is the
source-lens separation in units of $\theta_\e$.  Since typically
$\theta_\e\sim 0.5\,\mas$ and $\theta_*\sim 0.5\,\muas$, such effects
are expected for $A_\max\ga 1000$.  In the present case, we find that
the lens transited the source with an impact parameter of $\sim
\theta_*/3$, which permits a very accurate determination of
$\rho=8.50\pm 0.16\times 10^{-4}$.  
We evaluate the source radius $\theta_*$ by extending the method
of \citep{ob03262}, to three bands, $VIH$ \citep{ob06109b}.
  We obtain calibrated
$VI$ data from OGLE-II, and calibrated $H$ data by aligning 
SMARTS $H$-band to 2MASS \citep{2mass}.  We measure the
position of the clump and the source in 3 bands 
$(V,I,H)_{\rm cl} = (17.32,15.28,13.33)\pm(0.1,0.1,0.1)$,
$(V,I,H)_{\rm s} = (20.58,18.91,17.50)\pm(0.02,0.02,0.02)$,
We adopt
$(M_V,M_I,M_H)=(0.79,-0.25,-1.41)\pm(0.08,0.05,0.04)$ based on
\citet{alves02} adjusted for the $\alpha$-enhanced environment
of the bulge \citep{salaris02}.  We perform a Monte Carlo Markov
Chain (MCMC) optimization of a model with 4 parameters: (1) 
a \citet{cardelli89}
extinction law characterized by $R_V$, (2) the mean visual extinction
toward the clump $A_{V,\rm cl}$, (3) the mean distance to the clump, $R_0'$,
and (4) the extinction difference between the source and the clump
$\delta A_V$.  We assume (based on experience with high-resolution
spectra of bulge dwarfs) that $\delta A_V = 0 \pm 0.1$, and we demand
that the resulting $(V-I)_{0,\rm s}$ be consistent with that
predicted by \citet{bb88} from $(V-H)_{0,\rm s}$, with an error of 0.03 mag.
We then use the \citet{kervella04} $(V-H)$ color/surface-brightness
relation to derive $\theta_*=0.77\pm 0.03\,\muas$, where the error
also accounts for uncertainty in the model of the source flux 
(which affects all three bands in tandem).  As a check, we note
that this procedure yields $R_{VI} \equiv A_V/E(V-I) = 2.01 \pm 0.10$, 
a typical value for the bulge, and $R_0'= 7.8\pm 0.5\,$kpc, which is also
consistent with most estimates (keeping in mind that at 
$\ell = 2.37^\circ$, the bar is about 200 pc closer than the Galactic
center).

This result implies $\theta_\e = \theta_*/\rho =
0.91\pm 0.04\,\mas$.  Combining this with the measured Einstein
timescale, $t_\e=6.91\pm 0.13\,$days, yields a proper motion in the
Earth frame \citep{an02,gould04a} of $\mu_\geo = \theta_\e/t_\e =
48\pm 2\,\masyr$.

\subsection{Projected Einstein Radius From Terrestrial Parallax, $\bpi_\e$
\label{sec:pie}}

\begin{figure}
\plotone{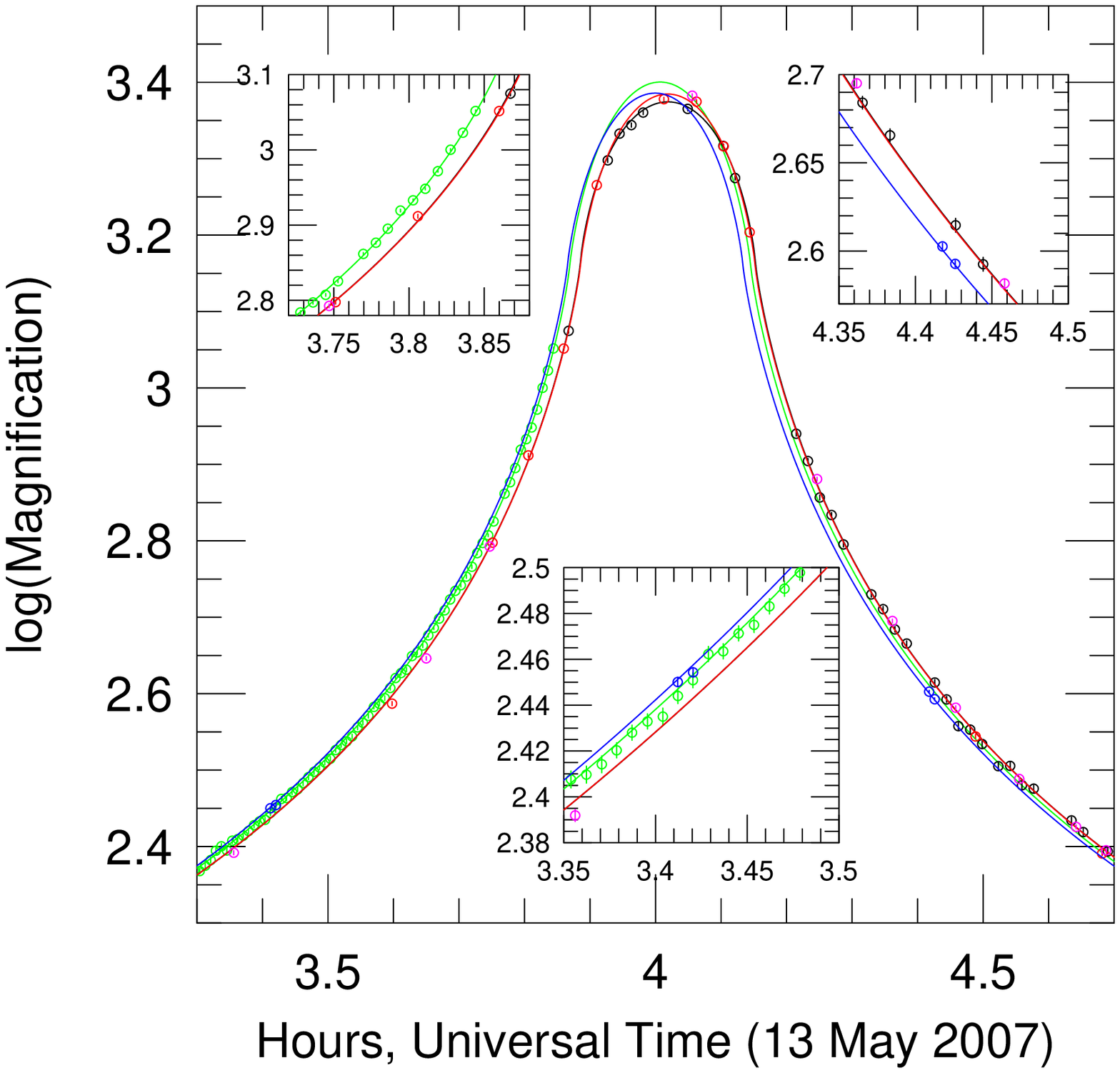}
\caption{Light curve of OGLE-2007-BLG-224 during 1.4 hours closest
to peak.  Observatories in South Africa (Bronberg: green),
Canaries (RoboNet LT: blue) and Chile 
(OGLE $I$: red, $\mu$FUN SMARTS $I$: magenta, $\mu$FUN SMARTS $H$: black)
see significantly (several percent) different magnifications due to their 
different positions on the Earth.  From these differences, one can
infer that $\tilde r_\e$ (the projected Einstein radius) 
is about 10,000 Earth radii.  Red and black curves (and points)
deviate slightly over the peak because of different limb-darkening
of the source in $I$ and $H$ bands.
}
\label{fig:lc}
\end{figure}

\begin{figure}
\plotone{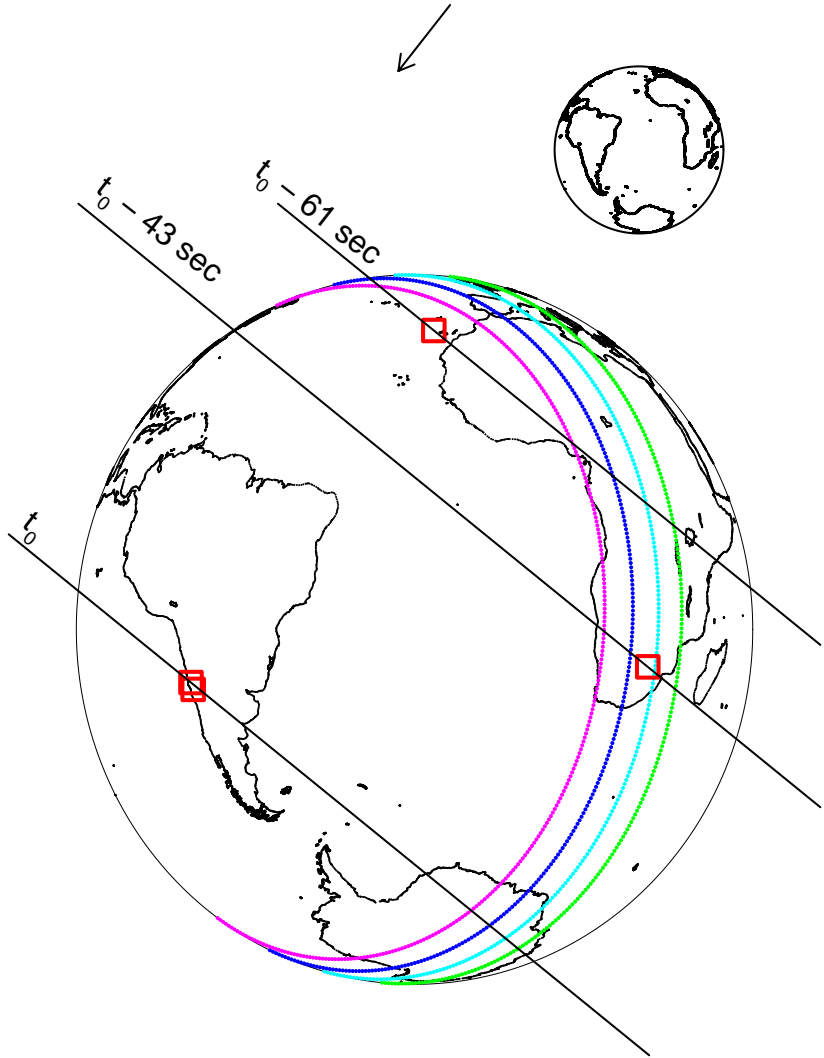}
\caption{Reconstruction of microlens fly-by.  Upper figure:
arrow defines locus of points from which lens and source appear
perfectly aligned, passing just 1 Earth diameter from the Earth's
surface.  Lower figure:  Earth as seen from OGLE-2007-BLG-224
at the peak of the event.  Red squares show the 4 observatories
(in Chile, South Africa, and the Canaries) that observed near
peak.  Black diagonal lines show contours of constant peak time,
with time offsets indicated between Chile and the other two locations.
Green curve shows sunrise, while cyan, blue, and magenta curves
show civil, nautical, and astronomical twilight.  As shown in
Fig.\ \ref{fig:lc}, South Africa observations stopped 10 minutes
before peak due to developing daylight.
}
\label{fig:earth}
\end{figure}

Second, and more dramatically, EMEs are subject to ``terrestrial parallax''
effects \citep{hardy95,holz96}.
If a microlensing event is observed from two different locations,
the source-lens relative trajectory will appear different: there will be
a different impact parameter $u_0$ and a different time of closest
approach $t_0$.  Since the projected Einstein radius $\tilde r_\e=\au/\pi_\e$
is typically of order AU, the second observer should typically be in solar
orbit to notice a significant effect, and indeed one such space-based
measurement has been made \citep{dong07}.   
However, the main way that microlens parallax
has been measured in the past is to take advantage of the moving platform
of the Earth, but usually the event must last a large fraction of
a year for this to work \citep{poindexter05}.  
For EMEs, the relevant spatial scale and
timescales are reduced by a factor $A_\max$.  For example, if the
projected velocity of the lens is $100\,\kms$ in the westward direction, then
$t_0$ will be about 80 seconds later in Chile than South Africa.  
Measuring the peak time of a normal microlensing event (which typically
lasts $t_\e\sim 30\,$days) to this precision is virtually impossible.  But 
since the peaking timescale of this event is roughly the source crossing time,
$t_*\equiv \rho\,t_\e=8.5\,$minutes, such measurements become quite 
practical.  Indeed
the event passed both South Africa and the Canaries about 1 minute
earlier than Chile, with both time differences accurate to a few seconds.
See Figure \ref{fig:lc}.
In practice, we simultaneously account for the difference in impact
parameters and peak times at all locations \citep{an02}, as well as the Earth's
orbital motion (which turns out be negligible), to measure both the
microlens parallax $\pi_\e = 1.97 \pm 0.13$ and the direction of 
lens-source relative motion, $52 \pm 5^\circ$
south of west.  See Figure \ref{fig:earth}.  The parallax measurement
can also be expressed in terms of the offsets,
${\bf \Delta t} \equiv (\Delta t_0,t_\e\Delta u_0)$, from which it is
ultimately derived:
$$
{\bf \Delta t}_{12}=(-18.1,54.0)s\pm (5.8,2.5)s
$$
$$
{\bf \Delta t}_{23}=(-43.1,-52.0)s\pm (2.6,7.0)s
$$
\begin{equation}
{\bf \Delta t}_{31}=(61.2,-2.0)s\pm (4.3,5.2)s,
\label{eqn:deltat}
\end{equation}
where 1 = South Africa, 2 = Canaries, 3 = Chile.

\subsection{Mass, Distance, and Transverse Velocity
\label{sec:mdpv}}

The determinations of $\pi_\e$ and $\theta_\e$ yield the mass and relative parallax
\begin{equation}
M = {\theta_\e\over \kappa\pi_\e} = 0.056\pm 0.004\,M_\odot
\quad \pi_\rel = \theta_\e\pi_\e = 1.78\pm 0.13\,\mas
\end{equation}
Assuming (as is consistent with its color and flux) that the source is
in the Galactic bulge $(\pi_s = 125\,\muas)$, the lens parallax and distance
are $\pi_l = \pi_\rel + \pi_s = 1.90\,\mas$, 
$D_l = \au/\pi_s = 525\pm 40\,\pc$.

The projected velocity of the lens in the Earth frame 
is $\tilde v_\geo = \mu_\geo(\au/\pi_\rel)= 127\,\kms$.  To find the
projected velocity in the frame of the Sun, we must remove the
motion of the Earth around the Sun ($23\,\kms$, almost due east).
We then find $\tilde v_\hel= 112\,\kms$ at $61^\circ$ south of west.
This means that the lens is moving against the direction of 
Galactic rotation, just $1^\circ$ out of the Galactic plane
(toward Galactic south).  Taking account of the motion of the Sun relative
to the local standard of rest \citep{dehnen98} as well as the small
mean motion (and its uncertainty) of the source, we find that
the lens is moving at $\tilde v_\hel= 113\pm 21\,\kms$ relative to the
Galactic disk at its location, almost directly counter to Galactic rotation.
This motion is quite consistent with the kinematics of the Galactic
thick disk, which has an asymmetric drift of $43\,\kms$ and a dispersion
of $49\,\kms$ in the direction of Galactic rotation \citep{casertano90}.
(It is also consistent with Galactic halo stars, but
with a probability more than 100 times smaller.)\ \ 
Moreover, since the inferred mass is below the threshold for
burning hydrogen, the lens is almost certainly a brown dwarf.  While
nearby brown dwarfs of this mass have been detected 
(e.g., \citealt{burgasser06,faherty09}), these are mostly
quite young and so still retain the heat generated during their collapse
from a cloud of gas.  The thick disk is of order 11 Gyrs old, and
hence brown dwarfs have had substantial time to cool.  Only those
very near the hydrogen burning limit are easy to see,
and then only if they are relatively nearby.

\section{Discussion
\label{sec:discuss}}

{\it Hubble Space Telescope (HST)} data were taken in $V$, $I$, $J$, 
and $H$ bands $\Delta t_1 =29$ days 
after peak when the source was magnified by a factor $A=1.005$, and again
almost exactly one year later, $\Delta t_2=1.08\,$ years.  At the
first epoch, the source and lens were virtually coincident, being
separated by only $\mu_\geo \Delta t_1\sim 4\,\mas$, i.e., $<0.1$
pixels, while at the second epoch they were separated by
$\mu_\hel \Delta t_2
\sim 47\,\mas\sim 1$ pixel, where $\mu_\hel =43\,\masyr$ is
the heliocentric proper motion.  These observations confirm that
the excess flux (above the source flux predicted from the microlensing
fit) is consistent with zero, but unfortunately not with very high
precision because an unrelated star lying 150 mas from the source
degrades the measurements.
 We discuss the prospect for future observations of the lens flux below.

We now ask, given a standard Galactic model \citep{han03,gould00}, how
likely it is that detailed information on a $t_\e=7\,$day (but otherwise
unconstrained) microlensing event toward this line of sight
$(l=2.37,\ b=-3.71)$ would turn out to imply a foreground ($D_l<4\,\kpc$)
thick-disk star rather than a star in the Galactic bulge?  The chance
is just 1/235.  
Hence, either we were extremely lucky to have found this object,
or brown dwarfs are more common in the thick disk than our standard
model for the mass function would predict.  Specifically, our model
assumes a relatively flat power law mass function
for low-mass stars and brown dwarfs with $dN/d\log\,M \propto M^{-0.3}$ from 
$M=0.7M_\odot$ down to $M=0.03\,M_\odot$.  This model is consistent
with results from observations of nearby thin disk brown dwarfs
\citep{pinfield06,metchev08}, and thus our detection would
imply that brown dwarfs must be an order of magnitude more
common in the thick disk than in the thin disk, in order that
the a priori probability of detecting this event be $\ga 10\%$.  This
possibility
should be considered seriously, since the prevalence of old brown
dwarfs is essentially unconstrained by any data, owing to their faintness.

In this light, we note that two other sets of investigators have concluded that
they must have been ``lucky'' unless old-population brown-dwarfs are more
common than generally assumed.  First, \citet{burgasser03} discovered a halo
brown dwarf, probably within 20 pc of the Sun, a volume that contains
only about 7 halo stars over the entire mass range from the hydrogen-burning
limit to a solar mass \citep{gould03}.  
Yet the survey in which this object was discovered
would only be sensitive to halo brown dwarfs in an extremely narrow mass
range just below this limit \citep{burgasser04}.  
Second, \citet{gaudi08}
analyzed a microlensing event of a nearby (1 kpc), bright ($V=11$) source
serendipitously discovered by an amateur astronomer hunting for comets,
finding that it was most likely a low-mass star or brown dwarf
moving at roughly $100\,\kms$.  Based on the low rate of such events
(and the non-systematic character of the search), they concluded that
they were ``probably lucky ... but perhaps not unreasonably so''
to have found this event.
See also \citet{fukui07}.
Finally, \citet{faherty09} found
14 high-velocity objects in a sample of 332 M, L, and T dwarfs,
which are consistent with thick-disk or halo kinematics.
Only one of these has $J-K<0$, which would be indicative of an old,
low-mass brown dwarf \citep{burrows01}, and with $v_{\rm tan}= 140\,\kms$
is most likely a halo brown dwarf, but could be in the high-velocity 
tail of the
thick-disk.  At absolute magnitude $M_J=15.9$, it must be very close
to the hydrogen-burning limit $M>0.07\,M_\odot$ if it is as old as
the thick disk and halo: $\ga 11\,$Gyrs.  As this object was found
within 10 pc, it is yet another ``lucky find'' 
(compare to, e.g., \citealt{burgasser04}).
While the evidence from each of these studies is marginal,
collectively they point to the need for closer investigation of the
frequency of old brown dwarfs.  

Brown dwarfs
are very blue in $H-K$ and somewhat red in $J-H$, so $H$ is the most
favorable band to observe them. 
In the heliocentric frame, the lens is moving away from the source
at $\mu_\hel=43\,\masyr$.
Hence, just 3 years after the event, the lens should be well removed from the
source in Keck telescope $H$-band images (given its $40\,\mas$ adaptive-optics
point spread function). Equivalent angular resolution with {\it HST} would 
require waiting four times longer, although it would profit from 
the much fainter sky as seen from space as well as its better-defined
point-spread function.  From \citet{burrows01} (Figs.\ 1,8,9,24) and
\citet{burrows97} (Fig.\ 23) one finds that 
the absolute magnitude of an 11 Gyr old, 
$M=0.056\,M_\odot$ brown dwarf is $M_H=17.0$, implying $H=25.7$ at
525 pc (assuming extinction $A_H=0.1$).  By comparison, the source
is $H_{\rm s} = 17.5$.  In principle, the lens could be
younger, in which case it would be brighter and so easier to detect.
However, such high-velocity young stars are extremely rare, and
there is no known mechanism for preferentially accelerating brown
dwarfs to this speed.  

If the lens is indeed in the thick disk, this would be
a challenging but not impossible observation, and would permit
a unique test of models of old brown dwarfs by comparison with
an object of known mass and well-constrained age.  The astrometry
associated with this measurement would yield  the lens-source relative proper
motion $\bmu_\hel$, which would test both the
{\it magnitude} of the microlens-based proper-motion and
the {\it direction} of the microlens parallax, thereby confirming
and refining (or contradicting) the microlens-based mass estimate
\citep{gould04b}.

Looking further ahead, the space-based {\it James Webb Space Telescope}
(launch currently scheduled for 2013) will be able to take a spectrum 
of this object, thereby testing more detailed atmospheric models.

\acknowledgments
We thank C.S. Kochanek, S.\ Poindexter and S.\ Kozlowski
for help with modeling the {\it HST} PSF.
We acknowledge the following support:
NSF AST-0757888 (AG,SD); NASA NNG04GL51G (DD,AG,RP);
Polish MNiSW N20303032/4275 (AU);
HST-GO-11311 (KS);
NSF AST-0206189, NASA NAF5-13042 (DPB);
Korea Science and Engineering Foundation grant 2009-008561 (CH)
Korea Research Foundation grant 2006-311-C00072 (B-GP)
Deutsche Forschungsgemeinschaft (CSB);
PPARC/STFC, EU FP6 programme ``ANGLES'' ({\L}W,NJR);
PPARC/STFC (RoboNet); 
Dill Faulkes Educational Trust (Faulkes Telescope North);
Grants JSPS18253002 and JSPS20340052 (MOA);
Marsden Fund of NZ(IAB, PCMY); Foundation for Research Science and
Technology of NZ.

\end{document}